# Entanglement distillation for *W* class states

Zhuo-Liang Cao[*], Ming Yang

Department of Physics, Anhui University, Hefei, 230039, PRChina

## Abstract

In this paper, we first propose a general entanglement distillation protocol for three-particle $W$ class state, which can concentrate the $W'$ state (non-maximally entangled $W$ state). The general protocol is mainly based on the unitary transformations on the auxiliary particles and the entangled particles, and a feasible physical scheme is suggested based on the cavity QED techniques. The protocol and scheme can be extended to the entanglement concentration of $N$-particle case.

[*] E-mail address: caoju@mars.ahu.edu.cn

# I Introduction

The non-locality property of entanglement illuminated by the Einstein, Podolsky and Rosen paradox [1] makes the entangled state a critical source for the quantum communication [2]. In recent years, there has been a rapid improvement in quantum communication [3-9]. But, if we want to realize a faithful teleportation of quantum state, the quantum channel must be a maximally entangled [2]. During the transmission, storage and processing, the entanglement of quantum state will unavoidably decrease because of noises. In practical case, entangled states distributed among distant locations are usually non-maximally entangled resulting from noises, which is resulting from the impossibility of increasing the entanglement of a system only by local operations and classical communications [10]. To achieve a faithful transmission of unknown states, we must, first, purify the noisy quantum channel. Bennett *et al* have proposed the first quantum purification scheme to purify some near perfectly entangled states out of a large supply of mixed entangled pairs using local operations [11]. But the implementation of the C-NOT gate used in Ref [11] is very difficult in experiment. We must find some physical processes to replace the theoretical logic gates. J.W. Pan *et al* have found a linear optical device, polarizing beam splitter, to take the role of the C-NOT gate [12]. Thereafter, some other theoretical and experimental schemes for the entanglement purification have been presented [13-20].

Vedral *et al* proposed the three basic steps needed by general entanglement purification protocols: (1) the local general measurements on the total system; (2) the classical communication; (3) postselection: the selection of the entangled pairs with higher purity conditioned on the measurement result of subsystems [21]. Along these steps, J. L. Romero *et al*, recently, proposed a physical scheme to purify the mixed entangled states of cavity modes [14]. We find that the purification of non-pure entangled states of bipartite system has been researched intensively. But there are few schemes for purifying the non-maximally entangled states of three particles in the literature, such as the non-maximally entangled $W$ states (in the article, we will use $W'$ to denotes it). In the previous article [22], we have concentrated the two-atom entangled state using cavity

QED techniques. Here we will discuss the distillation of $W'$ state.

W state is a special kind of entangled state. There is a more robust entanglement in it than in the GHZ state when one of the three particles was traced out [23]. When W state is used in the quantum communication, there will be some novel results [24]. So it is of practical significance to concentrate the $W'$ state.

The rest of the paper is outlined as follows: section II discusses the general distillation protocol for $W'$ state ( three-particle case and $N$-particle case), then a physical scheme realizing the general protocol will be discussed in section III, and the last part, section IV, is the conclusion.

## II The general distillation protocols for W′ state

The general form of $W$ state for the $N$ particles is:

$$|W_N\rangle = \frac{1}{\sqrt{N}}|N-1,1\rangle , \tag{1}$$

where $|N-1,1\rangle$ is the symmetric state involving $N$-1 zeros and 1 ones. Let $N$ =3, and we will get the W state:

$$|W_3\rangle = \frac{1}{\sqrt{3}}(|100\rangle + |010\rangle + |001\rangle) . \tag{2}$$

Here, we suppose that the three particles are initially prepared in the $W$ state in the form of (2), and, after interacting with the environments, it will evolve into the $W'$ state:

$$|W_3'\rangle = a|100\rangle + b|010\rangle + c|001\rangle . \tag{3}$$

Without loss of generality, we suppose that the coefficient a, b, c are complex numbers, $|a|^2 + |b|^2 + |c|^2 = 1$, $|a| \ge |b| \ge |c|$, and the subscripts denote the particles 1, 2 and 3. In fact, the $W$ state, interacting with different noises, probably evolves into a pure non-maximally entangled state or a mixed state. Here we will only consider the pure case. Assume that the three particles are shared by three distant users A, B and C. A has the access to particle 1, B has access to particle 2, and C has access to particle 3. Then the impure state can be concentrated by local operations and classical communication.

To extract the W state from $W'$ state, we will introduce two auxiliary particles, one in A's location, the other in B's location. The two auxiliary particles are initially prepared in the state $|0\rangle_{ai}$ ($i$=1, 2). Under the basis $\{|0\rangle_1|0\rangle_{a1}, |1\rangle_1|0\rangle_{a1}, |0\rangle_1|1\rangle_{a1}, |1\rangle_1|1\rangle_{a1}\}$, A will operate a joint unitary transformation on the particle 1 and the auxiliary particle $a1$. The unitary transformation, which is in the form:

$$U_1 = \begin{pmatrix} 1 & 0 & 0 & 0 \\ 0 & \frac{|c|}{a} & -\sqrt{1-\left|\frac{c}{a}\right|^2} & 0 \\ 0 & \sqrt{1-\left|\frac{c}{a}\right|^2} & \frac{|c|}{a^*} & 0 \\ 0 & 0 & 0 & 1 \end{pmatrix}, \tag{4}$$

will lead to the following evolution:

$$\begin{aligned}&\left(a|100\rangle_{123} + b|010\rangle_{123} + c|001\rangle_{123}\right)|0\rangle_{a1} \\ &\xrightarrow{U_1} \left(|c|\,|100\rangle_{123} + b|010\rangle_{123} + c|001\rangle_{123}\right)|0\rangle_{a1} + a\sqrt{1-|c/a|^2}\,|000\rangle_{123}|1\rangle_{a1}.\end{aligned} \tag{5}$$

At this moment, A will measure the auxiliary particle $a1$, and the state of the three particles will collapse into $|c|\,|100\rangle_{123} + b|010\rangle_{123} + c|001\rangle_{123}$ provided the fact that the auxiliary particle is in $|0\rangle_{a1}$. Then B will carry out another evolution by performing another joint unitary transformation on particle 2 and the auxiliary particle $a2$. This time, the transformation takes a new expression:

$$U_2 = \begin{pmatrix} 1 & 0 & 0 & 0 \\ 0 & \frac{|c|}{b} & -\sqrt{1-\left|\frac{c}{b}\right|^2} & 0 \\ 0 & \sqrt{1-\left|\frac{c}{b}\right|^2} & \frac{|c|}{b^*} & 0 \\ 0 & 0 & 0 & 1 \end{pmatrix}, \tag{6}$$

where the basis under which the transformation is constructed is $\{|0\rangle_2|0\rangle_{a2}, |1\rangle_2|0\rangle_{a2}, |0\rangle_2|1\rangle_{a2}, |1\rangle_2|1\rangle_{a2}\}$. Through the second transformation (6), the state $|c|\,|100\rangle_{123} + b|010\rangle_{123} + c|001\rangle_{123}$ will undergo the following evolution:

$$\left(|c|\,|100\rangle_{123}+b|010\rangle_{123}+c|001\rangle_{123}\right)|0\rangle_{a2}$$

$$\xrightarrow{U_2}\left(|c|\,|100\rangle_{123}+|c|\,|010\rangle_{123}+c|001\rangle_{123}\right)|0\rangle_{a2}+b\sqrt{1-|c/b|^2}\,|000\rangle_{123}|1\rangle_{a2}. \quad (7)$$

Then a measurement will be operated on the auxiliary particle *a*2. If the auxiliary particle is in the $|0\rangle_{a2}$ state, we have extracted the $W$ state from $W'$ state after C operated a rotational operation on particle 3:

$$|c|\times\sqrt{3}\times\frac{1}{\sqrt{3}}(|100\rangle_{123}+|010\rangle_{123}+|001\rangle_{123}), \quad (8)$$

and the success probability is :

$$P=3|c|^2. \quad (9)$$

During the protocol, if one of the auxiliary particles is in the $|1\rangle$ state, we could not extract a $W$ state from the $W'$ state, namely, the distillation fails.

As a straight extension of the above protocol, we will consider the concentration of N-particle $W'$ state, which is in the form:

$$|W'_N\rangle=c_1|10\cdots0\rangle+c_2|010\cdots0\rangle+\cdots+c_N|0\cdots01\rangle, \quad (10)$$

where $c_i(i=1,2,\cdots,N)$ are the complex coefficients of the state satisfying the condition $|c_1|+|c_2|+\cdots|c_N|=1$. There are $N$ users, each having access to one of the $N$ particles. We can assume that the $k$th user has access to the $k$th particle.

To realize the entanglement distillation of $N$-particle $W'$ state, there should be $N$-1 users each of whom must introduce an auxiliary particle in his location. At every location, the user must apply a unitary transformation on the two particles he has. Next we will consider the $k$th user as example. $ak$ is the very auxiliary particle in $k$th user's location initially prepared in $|0\rangle_{ak}$ state. Then the $k$th user will apply a unitary transformation which in the basis $\{|0\rangle_k|0\rangle_{ak},|1\rangle_k|0\rangle_{ak},|0\rangle_k|1\rangle_{ak},|1\rangle_k|1\rangle_{ak}\}$reads:

$$U_k = \begin{pmatrix} 1 & 0 & 0 & 0 \\ 0 & z_k & -\sqrt{1-|z_k|^2} & 0 \\ 0 & \sqrt{1-|z_k|^2} & z_k^* & 0 \\ 0 & 0 & 0 & 1 \end{pmatrix}, \tag{11}$$

where $z_k = \min(|c_1|, |c_2|, \cdots, |c_N|)/c_k$ (if $|c_j|$ is the $\min(|c_1|, |c_2|, \cdots, |c_N|)$, the $j$th user need not to operate this unitary transformation, that is why we only need $N$-1 auxiliary particles ). The transformation will lead to the evolution:

$$\begin{aligned} &\left(c_1|1\rangle_1|0\rangle_2\cdots|0\rangle_N + \cdots + c_k|0\rangle_1\cdots|0\rangle_{k-1}|1\rangle_k|0\rangle_{k+1}\cdots|0\rangle_N + \cdots + c_N|0\rangle_1\cdots|0\rangle_{N-1}|1\rangle_N\right)|0\rangle_{ak} \\ &\xrightarrow{U_k} \left[c_1|1\rangle_1|0\rangle_2\cdots|0\rangle_N + \cdots + c_{k-1}|0\rangle_1\cdots|0\rangle_{k-2}|1\rangle_{k-1}|0\rangle_k\cdots|0\rangle_N\right. \\ &\quad \left. + c_{k+1}|0\rangle_1\cdots|0\rangle_k|1\rangle_{k+1}|0\rangle_{k+2}\cdots|0\rangle_N + \cdots + c_N|0\rangle_1\cdots|0\rangle_{N-1}|1\rangle_N\right]|0\rangle_{ak} \\ &\quad + \min(|c_1|, |c_2|, \cdots, |c_N|)|0\rangle_1\cdots|0\rangle_{k-1}|1\rangle_k|0\rangle_{k+1}\cdots|0\rangle_N|0\rangle_{ak} \\ &\quad + c_k\sqrt{1-|z_k|^2}\,|0\rangle_1\cdots|0\rangle_N|1\rangle_{ak}. \end{aligned} \tag{12}$$

The $N$-1 users will apply the similar transformation on their own two particles. If after performing these transformations, all the $N$-1 users detect their ancillary particle in the $|0\rangle$, the $N$ particles are left in the $W$ state:

$$\min(|c_1|, |c_2|, \cdots, |c_N|)\sqrt{N}\frac{1}{\sqrt{N}}\left(|10\cdots0\rangle + |010\cdots0\rangle + \cdots + |0\cdots01\rangle\right), \tag{13}$$

provided the fact that the $j$th user apply a rotational operation on the $j$th particle. The success probability is:

$$P = N\min\left(|c_1|^2, |c_2|^2, \cdots, |c_N|^2\right) \tag{14}$$

After giving the general entanglement distillation protocols, we will present a feasible physical scheme, which can realize the general distillation via cavity QED techniques.

## III The physical scheme for the distillation of $W'$ state

In this section, we will use atoms as the carriers of the entanglement, cavities as the ancillary systems. Through the interaction between atom and cavity at every user's location, the non-maximally entangled *W* state can be concentrated.

Suppose the non-maximally entangled state of the N atoms is in the form:

$$\left|W_N'\right\rangle = c_1\left|eg\cdots g\right\rangle + c_2\left|geg\cdots g\right\rangle + \cdots + c_N\left|g\cdots ge\right\rangle, \tag{15}$$

where the parameters $c_i\left(i=1,2,\cdots,N\right)$ are just the same as that in Equation(10), and $|e\rangle$, $|g\rangle$ are the excited state and ground state of the atoms respectively. The preparation of this kind of states can be realized by non-linear interaction between atoms and other systems [25-27].

Assume that the *N* atoms have been distributed among *N* users at *N* distant locations. To concentrate the impure state $W'$, we must introduce *N*-1 ancillary systems, which are *N*-1 high fineness cavities at *N*-1 locations respectively. The cavities should be prepared in the vacuum state $\left|0\right\rangle$ initially, and there should be a detector of single photon corresponding to the cavity at each location.

Firstly, each of the *N*-1 users (just like the *N*-1 users in the general protocol) will send his atom through the corresponding cavity. Here we will consider the *k*th user for example. At *k*th user's location, the *k*th user will send the *k*th atom through the cavity, and the atom will interact with the cavity field. In the Jaynes-Cummings model for the interaction between a two-level atom and a single mode field, the Hamiltonian of the system can be expressed as:

$$\hat{H} = \omega a^+ a + \omega_0 S_z + \varepsilon\left(aS_+ + a^+ S_-\right), \tag{16}$$

where $\omega_0$ is the atomic transition frequency and $\omega$ is the cavity mode frequency, $a, a^+$ denote the annihilation and creation operators of the cavity mode, $S_+, S_-$ and $S_z$ are atomic operators, $S_+ = \left|e\right\rangle\left\langle g\right|$, $S_- = \left|g\right\rangle\left\langle e\right|$, $S_z = \frac{1}{2}\left(\left|e\right\rangle\left\langle e\right| - \left|g\right\rangle\left\langle g\right|\right)$; $\varepsilon$ is the coupling constant between atom and cavity mode. Here we can modulate the frequency of the

cavity mode so that the interaction is a resonant one.

After an interaction time $\Delta t_k$, the evolution of the state of the total system is as follows:

$$\left(c_1|e\rangle_1|g\rangle_2\cdots|g\rangle_N+\cdots+c_k|g\rangle_1\cdots|g\rangle_{k-1}|e\rangle_k|g\rangle_{k+1}\cdots|g\rangle_N+\cdots+c_N|g\rangle_1\cdots|g\rangle_{N-1}|e\rangle_N\right)|0\rangle_{ak}$$

$$\xrightarrow{U_k}\exp\left(\frac{i\omega\Delta t_k}{2}\right)\left[c_1|e\rangle_1|g\rangle_2\cdots|g\rangle_N+\cdots+c_{k-1}|g\rangle_1\cdots|g\rangle_{k-2}|e\rangle_{k-1}|g\rangle_k\cdots|g\rangle_N\right.$$

$$\left.+c_{k+1}|g\rangle_1\cdots|g\rangle_k|e\rangle_{k+1}|g\rangle_{k+2}\cdots|g\rangle_N+\cdots+c_N|g\rangle_1\cdots|g\rangle_{N-1}|e\rangle_N\right]|0\rangle_{ak}$$

$$+\exp\left(-\frac{i\omega\Delta t_k}{2}\right)c_k\cos\varepsilon t|g\rangle_1\cdots|g\rangle_{k-1}|e\rangle_k|g\rangle_{k+1}\cdots|g\rangle_N|0\rangle_{ak}$$

$$-i\exp\left(-\frac{i\omega\Delta t_k}{2}\right)c_k\sin\varepsilon t|g\rangle_1\cdots|g\rangle_N|1\rangle_{ak}. \quad (17)$$

If he selects the optimal interaction times:

$$\Delta t_k=\frac{1}{\varepsilon}\cos^{-1}\frac{\min\left(|c_1|,|c_2|,\cdots,|c_N|\right)}{|c_k|}, \quad (18)$$

we can realize:

$$|c_k|\cos\varepsilon\,\Delta t_k=\min\left(|c_1|,|c_2|,\cdots,|c_N|\right). \quad (19)$$

Then the $k$th user will detect the cavity. If all users detect their cavity in the vacuum state $|0\rangle$, the $N$ atoms are left in the $W$ state, and the success probability for this is:

$$P=N\min\left(|c_1|^2,|c_2|^2,\cdots,|c_N|^2\right). \quad (20)$$

Here we have disregard the phase factor, because we can modulate it, easily, by sending the atom through a classical Ramsey zone.

That is to say, the success probability is only determined by the smallest coefficient of the superposition state to be concentrated. After distillation, the $N$ atoms are in the $W$ state, which is a more robust resource in the quantum communication [24].

# IV Conclusion

In conclusion, we have presented a general entanglement distillation protocol for non-maximally entangled $W$ state, and found a detailed physical scheme to realize the general protocol via cavity QED. In the protocol, we introduced two auxiliary particles for the three-particle case, and realized the distillation of $W$ state by coupling two of the three particles to the corresponding auxiliary systems and thereafter performing a postselection depending on the result of detection performed on the latter. As an extension, we considered the $N$ particle case in the similar way. Then we presented an effective physical scheme to realize the distillation of the $N$-1 non-maximally entangled $W$ state for atomic system, where we used atoms as the carriers of the entanglement, cavities as the ancillary systems. The success probabilities are all dependent on the smaller coefficient of the superposition of the $W'$ state. Here we only discussed the pure state cases. For the more general case, purification of mixed state, we will discuss in the future publication.

## Acknowledgements

This work is supported by the Natural Science Foundation of Anhui Province under Grant No: 03042401 and the Natural Science Foundation of the Education Department of Anhui Province under Grant No: 2002kj026, also by the fund of the Core Teacher of Ministry of National Education under Grant No: 200065.